%Paper: hep-ph/9301220
%From: FHALON@WISWIC.WEIZMANN.AC.IL
%Date: Fri, 8 Jan 1993 10:59:18 GMT
%Date (revised): Fri, 8 Jan 1993 11:47:49 GMT

%
% Printing instructions:
%       This paper needs the macro packages phyzzx.tex and tables.tex.
%
%
\input phyzzx
\hsize=5.5in
\vsize=8.70in
\tolerance=2000
\sequentialequations
\def\rl{\rightline}

\def\t1{{\tilde 1}}

\def\AEF{A.E. Faraggi}
\def\DVN{D. V. Nanopoulos}

\def\NPB#1#2#3{Nucl. Phys. B{\bf#1} (19#2) #3}
\def\PLB#1#2#3{Phys. Lett. B{\bf#1} (19#2) #3}

\REF\FFF{I. Antoniadis and C. Bachas,
\NPB{298}{88}{586}; H. Kawai, D.C. Lewellen, and S.H.-H. Tye,
\NPB{288}{87}{1}.}
\REF\DHVW{L. Dixon, J.A. Harvey, C. Vafa and E. Witten, \NPB{261}{85}{678},
\NPB{274}{86}{285}.}
\REF\REVAMP{I. Antoniadis, J. Ellis,
J. Hagelin, and \DVN, \PLB{231}{89}{65}.}
\REF\FNY{\AEF, D.V. Nanopoulos and K. Yuan, \NPB{335}{90}{437}.}
\REF\EU{\AEF, \PLB{278}{92}{131} .}
\REF\TOP{\AEF, \PLB{274}{92}{47}.}
\REF\SLM{\AEF, WIS--91/83/NOV--PH, \AEF, WIS--92/16/FEB--PH ( To appear in
Nucl.
Phys. B.). }
\REF\KLN{S. Kalara, J. Lopez and D.V. Nanopoulos,
\PLB{245}{91}{421}; \NPB{353}{91}{650}.}
\REF\DSW{M. Dine, N. Seiberg and E. Witten, \NPB{289}{87}{585}.}
\REF\NRT{\AEF, WIS--92/48/JUN--PH.}
\REF\GCU{\AEF, WIS--92/17/FEB--PH.}

%\magnification=1200
\singlespace
\rl{WIS--92/103/DEC--PH}
%\rl{\today}
%\rl{T}
\normalspace
\bigskip
\titlestyle{\bf{Construction of Realistic superstring Standard--like
Models}
{\footnote*{talk presented at the TEXAS\\PASCHOS conference, Berkeley
CA, December 13--18 1992}}
}
\bigskip
\centerline{Alon E. Faraggi
%{\footnote*{e--mail address: fhalon@weizmann.bitnet}}
}
\smallskip
\centerline {\it Department of Physics, Weizmann Institute of Science}
\centerline {\it Rehovot 76100, Israel}
%\titlestyle{ABSTRACT}
\bigskip

I discuss the construction of realistic superstring standard--like models
in the four dimensional free fermionic formulation. I discuss
proton lifetime constraints on
superstring models. I discuss the massless spectrum of the
superstring standard--like models, the
texture of fermion mass matrices in these models and argue that the
realistic features of these models are due to the underlying $Z_2\times
Z_2$ orbifold compactification.

\singlespace
\normalspace
\pagenumber 1

The prospects for grand unification and especially supersymmetric
grand unification seem very promising in view of LEP precision data.
However, point field theories are in general incomplete.
There are many problems whose solution cannot be found in the context
of point quantum field theories. First, there are too many free parameters
and the choice of the gauge group is arbitrary. Second, how does nature
choose to have only three generations and the mass hierarchy among these
generations, especially the top mass hierarchy relative to the lighter
quarks and leptons. Finally, quantum gravity and point quantum field theories
are incompatible.

In the context of unified theories the resolution of these problems must
come from the fundamental gravitational--geometrical theory at the Planck
scale. Superstring theory is the most developed Planck scale theory to
date. Moreover, it is the only theory we know that can consistently
unify gravity with the gauge interactions. Therefore, an extremely
important task is to try to connect superstring theory with the
standard model of the gauge interactions.

Two approaches can be pursued to connect superstring theory with the
standard model. One is through a GUT model at an intermediate
energy scale. However, the only GUT superstring models constructed to date
have level one Kac--Moody algebras and those do not have adjoint
representations. Therefore, one has to construct superstring GUT models
that do not use adjoint representations to further break the symmetry
to the standard model. The second possibility is to derive the
standard model directly from superstring theory.

In this talk I focus on the second possibility. This focus is motivated
by consideration of proton decay. In the most general supersymmetric
standard model the dimension four operators,
${\eta_1}{u_{L}^C}{d_{L}^C}{d_{L}^C}+{\eta_2}{d_{L}^C}QL$,
mediate rapid proton decay. In the minimal supersymmetric standard model
one imposes a discrete symmetry, R parity, that forbids these terms.
In the context of superstring theories discrete symmetries are
usually not present. If $B-L$ is gauged as in $SO(10)$, these dimensions
four operators are forbidden by gauge invariance. However they may still
be induced from the nonrenormalizable terms,
$${\eta_1}({u_{L}^C}{d_{L}^C}{d_{L}^C}N_L^C)\Phi+
{\eta_2}({d_{L}^C}QLN_L^C)\Phi,\eqno(1)$$
where $\Phi$ is a string of $SO(10)$ singlets that fixes the string
selection rules and gets a VEV of $O(m_{pl})$. $N_L^C$ is the standard model
singlet in the $16$ of $SO(10)$. It is seen that the ratio
${\langle{N_L^c}\rangle}/{M_{pl}}$ controls the rate
of proton decay. Consequently, the VEV ${\langle{N_L^c}\rangle}$ has to be
suppressed.
In superstring GUT models, ${\langle{N_L^c}\rangle}$
must be used to break the GUT symmetry as there are no adjoint representations.
Thus, it is concluded that we must obtain the standard model directly
from superstring theory.

The standard--like models are constructed in the free fermionic
formulation. In the free fermionic formulation$^\FFF$ of the heterotic string
in four dimensions all the world--sheet
degrees of freedom  required to cancel
the conformal anomaly are represented  in terms of free fermions
propagating on the string world--sheet.
Under parallel transport around a noncontractible loop the
fermionic states pick up a phase. A model in this construction
is defined by a set of  basis vectors of boundary conditions for all
world--sheet fermions. These basis vectors are constrained by the string
consistency requirements (e.g. modular invariance) and
completely determine the vacuum structure of
the model. The physical spectrum is obtained by applying the generalized
GSO projections. The low energy effective field theory is obtained
by S--matrix elements between external states. The Yukawa couplings
and higher order nonrenormalizable terms in the superpotential
are obtained by calculating corralators between vertex operators.
For a corralator to be nonvanishing all the symmetries of the model must
be conserved. Thus, the boundary condition vectors determine the
phenomenology of the models.

The first five vectors (including the vector {\bf 1}) in the basis
consist of the NAHE{\footnote*{This set was first
constructed by Nanopoulos, Antoniadis, Hagelin and Ellis  (NAHE)
in the construction
of  the flipped $SU(5)$.  {\it nahe}=pretty, in
Hebrew.}} set
$$\eqalignno{S&=({\underbrace{1,\cdots,1}_{{\psi^\mu},
{\chi^{1..6}}}},0,\cdots,0
\vert 0,\cdots,0).&(2a)\cr
b_1&=({\underbrace{1,\cdots\cdots\cdots,1}_
{{\psi^\mu},{\chi^{12}},y^{3,...,6},{\bar y}^{3,...,6}}},0,\cdots,0\vert
{\underbrace{1,\cdots,1}_{{\bar\psi}^{1,...,5},
{\bar\eta}^1}},0,\cdots,0).&(2b)\cr
b_2&=({\underbrace{1,\cdots\cdots\cdots\cdots\cdots,1}_
{{\psi^\mu},{\chi^{34}},{y^{1,2}},
{\omega^{5,6}},{{\bar y}^{1,2}}{{\bar\omega}^{5,6}}}}
,0,\cdots,0\vert {\underbrace{1,\cdots,1}_{{{\bar\psi}^{1,...,5}},
{\bar\eta}^2}}
,0,\cdots,0).&(2c)\cr
b_3&=({\underbrace{1,\cdots\cdots\cdots\cdots\cdots,1}_
{{\psi^\mu},{\chi^{56}},{\omega^{1,\cdots,4}},
{{\bar\omega}^{1,\cdots,4}}}},0,\cdots,0
\vert {\underbrace{1,\cdots,1}_{{\bar\psi}^{1,...,5},
{\bar\eta}^3}},0,\cdots,0).&(2d)\cr}$$
with the choice of generalized GSO projections
$c\left(\matrix{b_i\cr
                                    b_j\cr}\right)=
c\left(\matrix{b_i\cr
                                    S\cr}\right)=
c\left(\matrix{1\cr
                                    1\cr}\right)=-1,$
%\eqno(2)
and the others given by modular invariance.

The gauge group after the NAHE set is $SO(10)\times
E_8\times SO(6)^3$ with $N=1$ space--time supersymmetry,
and 48 spinorial $16$ of $SO(10)$, sixteen from each sector
$b_1$, $b_2$ and $b_3$. The NAHE set divides the internal world--sheet
fermions in the following way: ${\bar\phi}^{1,\cdots,8}$ generate the
hidden $E_8$ gauge group, ${\bar\psi}^{1,\cdots,5}$ generate the $SO(10)$
gauge group, and $\{{\bar y}^{3,\cdots,6},{\bar\eta}^1\}$,
$\{{\bar y}^1,{\bar y}^2,{\bar\omega}^5,{\bar\omega}^6,{\bar\eta}^2\}$,
$\{{\bar\omega}^{1,\cdots,4},{\bar\eta}^3\}$ generate the three horizontal
$SO(6)^3$ symmetries. The left--moving $\{y,\omega\}$ states are divided
to $\{{y}^{3,\cdots,6}\}$,
$\{{y}^1,{y}^2,{\omega}^5,{\omega}^6\}$,
$\{{\omega}^{1,\cdots,4}\}$ and $\chi^{12}$, $\chi^{34}$, $\chi^{56}$
generate the left--moving $N=2$ world--sheet supersymmetry.

The internal fermionic states $\{y,\omega\vert{\bar y},{\bar\omega}\}$
correspond to the six left--moving and six right--moving compactified
dimensions in a geometric formulation. This correspondence is illustrated
by adding the vector
$X=(0_L\vert{{\bar\psi}^{1,\cdots,5}},
{{\bar\eta}^{1,2,3}},0,\cdots,0)$
to the NAHE set, which extends the gauge symmetry to
$E_6\times U(1)^2\times E_8\times SO(4)^3$ with $N=1$ supersymmetry
and twentyfour chiral $27$ of $E_6$. The same model is generated in the
orbifold language$^\DHVW$
by moding out an $SO(12)$ lattice by a $Z_2\times{Z_2}$
discrete symmetry with standard embedding. In the construction of
the standard--like models beyond the NAHE set, the assignment
of boundary conditions to the set of internal fermions
$\{y,\omega\vert{\bar y},{\bar\omega}\}$ determines many of the
properties of the low energy spectrum, such as the number of
generations, the presence of Higgs doublets, Yukawa couplings, etc.

The standard--like models are constructed by adding three additional
vectors to the NAHE set$^{\FNY,\EU,\TOP,\SLM}$.
One example is presented in the table, where only the boundary conditions
of the ``compactified space" are shown. In the gauge sector
$\alpha,\beta\{{{\bar\psi}^{1,\cdots,5}},
{{\bar\eta}^{1,2,3}},{\bar\phi}^{1,\cdots,8}\}=\{1^3,0^5,1^4,0^4\}$
and $\gamma\{{{\bar\psi}^{1,\cdots,5}},
{{\bar\eta}^{1,2,3}},{\bar\phi}^{1,\cdots,8}\}=\{{1\over2}^9,0,1^2,
{1\over2}^3,0\}$ break the symmetry to $SU(3)\times SU(2)\times
U(1)_{B-L}\times U(1)_{T_{3_R}}\times SU(5)_h\times SU(3)_h\times U(1)^2$.
The choice of generalized GSO coefficients is:
${c\left(\matrix{b_j\cr
                                    \alpha,\beta,\gamma\cr}\right)=
-c\left(\matrix{\alpha\cr
                                    1\cr}\right)=
c\left(\matrix{\alpha\cr
                                    \beta\cr}\right)=
-c\left(\matrix{\beta\cr
                                    1\cr}\right)}=$
${c\left(\matrix{\gamma\cr
                                    1,\alpha\cr}\right)=
-c\left(\matrix{\gamma\cr
                                    \beta\cr}\right)=
-1}$ (j=1,2,3), with the others specified by modular invariance and
space--time supersymmetry.
Three additional vectors are needed to reduce the
number of generations to one generation from each sector $b_1$, $b_2$
and $b_3$. Each generation has horizontal symmetries that constrain
the allowed interactions. Each generation has two gauged $U(1)$
symmetries $U(1)_{R_j}$ and $U(1)_{R_{j+3}}$. For every right--moving
$U(1)$ symmetry there is a corresponding left--moving global $U(1)$ symmetry
$U(1)_{L_j}$ and $U(1)_{L_{j+3}}$. Finally, each generation has two Ising
model operators that are obtained by pairing a left--moving real fermion
with a right--moving real fermion.

\input tables.tex
%\special{landscape}
%\hoffset=1.25truein
%\nopagenumbers
%\magnification=1000
%\font\normalroman=cmr10
%\font\style=cmr7
%\style
\tolerance=1200

%\fontdimen12\fivesy=0pt

%\textfont0=\sevenrm
%\scriptfont0=\fiverm
%\textfont1=\seveni
%\scriptfont1=\fivei
%\textfont2=\sevensy
%\scriptfont2=\fivesy

\smallskip
{\hfill
{\begintable
\  \ \|\
${y^3y^6}$,  ${y^4{\bar y}^4}$, ${y^5{\bar y}^5}$,
${{\bar y}^3{\bar y}^6}$
\ \|\ ${y^1\omega^6}$,  ${y^2{\bar y}^2}$,
${\omega^5{\bar\omega}^5}$,
${{\bar y}^1{\bar\omega}^6}$
\ \|\ ${\omega^1{\omega}^3}$,  ${\omega^2{\bar\omega}^2}$,
${\omega^4{\bar\omega}^4}$,  ${{\bar\omega}^1{\bar\omega}^3}$  \crthick
$\alpha$ \|
1, ~~~0, ~~~~0, ~~~~0 \|
0, ~~~0, ~~~~1, ~~~~1 \|
0, ~~~0, ~~~~1, ~~~~1 \nr
$\beta$ \|
0, ~~~0, ~~~~1, ~~~~1 \|
1, ~~~0, ~~~~0, ~~~~0 \|
0, ~~~1, ~~~~0, ~~~~1 \nr
$\gamma$ \|
0, ~~~1, ~~~~0, ~~~~1 \|\
0, ~~~1, ~~~~0, ~~~~1 \|
1, ~~~0, ~~~~0, ~~~~0  \endtable}
\hfill}
\smallskip
\parindent=0pt
\hangindent=39pt\hangafter=1
%\normalroman

%\vfill
%\eject
%\vskip 5cm

Higgs doublets in the standard--like models are obtained from two
distinct sectors. The first are obtained from the Neveu--Schwarz sector,
which produce three pairs of electroweak doublets. Each pair can couple
at tree level only to the states from the sector $b_j$. There is
stringy doublet--triplet spilting mechanism that projects out the
color triplets and leaves the electroweak doublets in the spectrum.
Thus, the superstring standard--like models resolve the GUT hierarchy
problem. The second type of Higgs doublets are obtained from
the vector combination $b_1+b_2+\alpha+\beta$. The states in this
sector are obtained by acting on the vacuum with a single fermionic
oscillator and transform only under the observable sector.

The cubic level Yukawa couplings for the quarks and leptons are
determined by the boundary conditions in the vector $\gamma$
according to the following rule
$$\eqalignno{\Delta_j&=
\vert\gamma(U(1)_{\ell_{j+3}})-\gamma(U(1)_{r_{j+3}})\vert=0,1
{\hskip 1cm}(j=1,2,3)&(3a)\cr
\Delta_j&=0\rightarrow d_jQ_jh_j+e_jL_jh_j;{\hskip .2cm}
\Delta_j=1\rightarrow u_jQ_j{\bar h}_j+N_jL_j{\bar h}_j,&(3b,c)\cr}$$ where
$\gamma(U(1)_{R_{j+3}})$, $\gamma(U(1)_{\ell_{j+3}})$ are the boundary
conditions of the world--sheet fermionic currents that generate the
$U(1)_{R_{j+3}}$, $U(1)_{\ell_{j+3}}$ symmetries.

The superstring standard--like models contain an anomalous $U(1)$
gauge symmetry. The anomalous $U(1)$ generates a Fayet--Iliopoulos term
by the VEV of the dilaton field that breaks supersymmetry and destabelizes
the vacuum. Supersymmetry is restored by giving VEVs to standard model
singlets in the massless spectrum of the superstring models. However,
as the charge of these singlets must have $Q_A<0$ to cancel the anomalous
$U(1)$ D--term equation, in  many models a phenomenologically realistic
solution does not exist. In fact a very restricted class of standard--like
models with $\Delta_j=1$ for $j=1,2,3$, were found to admit a solution
to the F and D flatness constraints. Consequently, the only models that
were found to admit a solution are models which have tree level
Yukawa couplings only for $+{2\over3}$ charged quarks.

This result suggests an explanation for the top quark mass hierarchy
relative to the lighter quarks and leptons. At the cubic level only the
top quark gets a mass term and the mass terms for the lighter
quarks and leptons are obtained from nonrenormalizable terms.
To study this scenario we have to examine the nonrenormalizable
contributions to the doublet Higgs mass matrix and to the fermion mass
matrices.

At the cubic level there are two pairs of electroweak doublets.
At the nonrenormalizable level one additional pair receives a superheavy
mass and one pair remain light to give masses to the fermions at
the electroweak scale. Requiring F--flatness imposes that the light
Higgs representations are ${\bar h}_1$ or ${\bar h}_2$ and $h_{45}$.

The nonrenormalizable fermion mass terms of order N are of the form
$cgf_if_jh\phi^{^{N-3}}$ or
$f_if_j{\bar h}\phi^{^{N-3}}$, where $c$ is a
calculable coefficient, $g$ is the gauge coupling at the unification
scale,  $f_i$, $f_j$ are the fermions from
the sectors $b_1$, $b_2$ and $b_3$, $h$ and ${\bar h}$ are the light
Higgs doublets, and $\phi^{N-3}$ is a string of standard model singlets
that get a VEV and produce a suppression factor
${({{\langle\phi\rangle}\over{M}})^{^{N-3}}}$ relative to the cubic
level terms. Several scales contribute to the generalized VEVs. The
leading one is the scale of VEVs that are used to cancel the anomalous
D--term equation. The next scale is generated by Hidden sector
condensates. Finally, there is a scale which is related to the breaking
of $U(1)_{Z^\prime}$, $\Lambda_{Z^\prime}$. Examination of the higher
order nonrenormalizable terms reveals that $\Lambda_{Z^\prime}$ has
to be suppressed relative to the other two scales.

At the cubic level only the top quark gets a nonvanishing mass term.
Therefore only the top quark mass is characterized by the electroweak
scale. The remaining quarks and leptons obtain their mass terms from
nonrenormalizable terms. The states from the sectors $b_1$ and $b_2$
produce the remaining states for the two heaviest generations.
Their mass terms are suppressed by singlet VEVs that are used in the
cancellation of the anomalous $U(1)$ D--term equation. The diagonal
mass terms for the states from $b_3$ can only be generated by
VEVs that break $U(1)_{Z^\prime}$. Therefore, the diagonal mass terms
for the lightest generation states are suppressed. Similarly, terms that
mix between different generations can only be generated by hidden sector
states from the sectors $b_j+2\gamma$. If the hidden sector states
obtain a VEV in the cancellation of the anomalous $U(1)$ D--term
equation, Cabbibo mixing of the correct order of magnitude can be
generated. To summarize, the texture of the fermion mass matrices in
the superstring standard--like models is of the following form,
$${M_U\sim\left(\matrix{\epsilon,a,b\cr
                    {\tilde a},A,c \cr
                    {\tilde b},{\tilde c},\lambda_t\cr}\right);{\hskip .2cm}
M_D\sim\left(\matrix{\epsilon,d,e\cr
                    {\tilde d},B,f \cr
                    {\tilde e},{\tilde f},C\cr}\right);{\hskip .2cm}
M_E\sim\left(\matrix{\epsilon,g,h\cr
                    {\tilde g},D,i \cr
                    {\tilde h},{\tilde i},E\cr}\right)},$$
where $\epsilon\sim({{\Lambda_{Z^\prime}}\over{M}})^2$.
The diagonal terms in capital letters represent leading
terms that are suppresses by singlet VEVs, and
$\lambda_t=O(1)$. The mixing terms are generated by hidden sector states
from the sectors $b_j+2\gamma$ and are represented by small letters. They
are proportional to $({{\langle{TT}\rangle}\over{M}^2})$.

Next, I turn to the problem of gauge coupling
unification in the superstring standard--like models. The problem
is that while LEP results indicate that the gauge coupling in the minimal
supersymmetric standard model unify at $10^{16}GeV$, superstring
theory predicts that the unification scale is at $10^{18}GeV$.
The superstring standard--like models may resolve this
problem due to the existence of color triplets and electroweak doublets
from exotic sectors. These exotic states carry fractional charges and
do not fit into standard $SO(10)$ representations.
The standard--like models predict
$\sin^2\theta_W={3\over8}$ at the unification scale due to the
embedding of the weak hypercharge in $SO(10)$. In Ref. (\GCU),
I showed that provided that the additional exotic color triplets and
electroweak doublets exist at the appropriate scales, the
scale of gauge coupling unification is pushed to $10^{18}GeV$, with the
correct value of $\sin^2\theta_W$at low energies.

To conclude, the superstring standard--like models represent the
most developed superstring models to date. They contain in their
massless spectrum all the necessary states to obtain realistic
phenomenology. They resolve the problems of proton decay through
dimension four and five operators that are endemic to other
superstring and GUT models. The free fermionic standard-like models
provide a highly constrained and phenomenologically realistic laboratory
to study how the Planck scale affects low energy physics.

\bigskip
%\centerline{\bf Acknowledgments}
%I would like to thank Lance Dixon for useful discussions.
%This work is supported in part by a Feinberg School Fellowship.

\baselineskip=15pt
\refout

\vfill
\eject

\end

\input tables.tex
%\special{landscape}
%\hoffset=1.25truein
%\nopagenumbers
%\magnification=1000
%\font\normalroman=cmr10
%\font\style=cmr7
%\style
\tolerance=1200

%\fontdimen12\fivesy=0pt

%\textfont0=\sevenrm
%\scriptfont0=\fiverm
%\textfont1=\seveni
%\scriptfont1=\fivei
%\textfont2=\sevensy
%\scriptfont2=\fivesy

\smallskip
{\hfill
{\begintable
\  \ \|\
${y^3y^6}$,  ${y^4{\bar y}^4}$, ${y^5{\bar y}^5}$,
${{\bar y}^3{\bar y}^6}$
\ \|\ ${y^1\omega^6}$,  ${y^2{\bar y}^2}$,
${\omega^5{\bar\omega}^5}$,
${{\bar y}^1{\bar\omega}^6}$
\ \|\ ${\omega^1{\omega}^3}$,  ${\omega^2{\bar\omega}^2}$,
${\omega^4{\bar\omega}^4}$,  ${{\bar\omega}^1{\bar\omega}^3}$  \crthick
$\alpha$ \|
1, ~~~0, ~~~~0, ~~~~0 \|
0, ~~~0, ~~~~1, ~~~~1 \|
0, ~~~0, ~~~~1, ~~~~1 \nr
$\beta$ \|
0, ~~~0, ~~~~1, ~~~~1 \|
1, ~~~0, ~~~~0, ~~~~0 \|
0, ~~~1, ~~~~0, ~~~~1 \nr
$\gamma$ \|
0, ~~~1, ~~~~0, ~~~~1 \|\
0, ~~~1, ~~~~0, ~~~~1 \|
1, ~~~0, ~~~~0, ~~~~0  \endtable}
\hfill}
\smallskip
\parindent=0pt
\hangindent=39pt\hangafter=1
%\normalroman
\baselineskip=18pt

{{\it Table 2.}
%A three generations ${SU(3)\times SU(2)\times U(1)^2}$
%model.
%Trilevel Yukawa couplings are obtained only for
%${+{2\over3}}$ charged quarks.}
\bigskip
%\magnification=1200